# Comment on the paper "Third-harmonic generation investigated by a short-range bottomless exponential potential well" by M. Hu, K. Guo, Q. Yu, Z. Zhang [Superlattices and Microstructures, 122 (2018) 538-547]

**A.M. Ishkhanyan[1,2] and G.G. Demirkhanyan[2,3]**

[1]Russian-Armenian University, Yerevan, 0051 Armenia
[2]Institute for Physical Research, Ashtarak, 0203 Armenia
[3]Armenian State Pedagogical University, Yerevan, 0010 Armenia

We have discovered several severe errors in the recent paper M. Hu, K. Guo, Q. Yu, Z. Zhang [Superlattices and Microstructures, **122** (2018) 538-547]. Specifically, we demonstrate that both the solution of the Schrödinger equation and the bound-state wave functions used in the paper are incorrect.

In the recent paper [1], the authors discuss the third-harmonic generation investigated by a short-range bottomless exponential potential well.

Unfortunately, the paper contains several severe mistakes.

**1.** The short-range bottomless exponential potential the authors discuss is

$$V = \frac{V_0}{\sqrt{1-e^{-y/\delta}}} - V_0. \qquad (1)$$

By the change of the variables $\psi = u(x)\varphi(x)$ with $x = x(y) = \sqrt{1-e^{-y/\delta}}$, the authors reduce the Schrödinger equation

$$\frac{d^2\psi}{dy^2} + (E - V(y))\frac{2m}{\hbar^2}\psi = 0 \qquad (2)$$

to the Heun equation

$$\frac{d^2u}{dx^2} + \frac{du}{dx}\left(\frac{a}{x+1} + \frac{b}{x} + \frac{c}{x-1}\right) - u\frac{k - \beta\gamma x}{x(x+1)(x-1)} = 0. \qquad (3)$$

The authors claim that here $a = -1$. However, one checks that the parameter $b$, not $a$, is equal to minus one. This is an important point which plays a key role in constructing the solution of the equation (see below).

Next, the authors claim that the solution of the Heun equation (3) is given as

$$u(x) = \left(c_1 H_\rho(z) + {}_1F_1\left(-\frac{\rho}{2}; \frac{1}{2}; z^2\right)c_2\right)e^{-\sqrt{2\rho}z}, \qquad (4)$$

where $H_\rho$ is the Hermite function, ${}_1F_1$ is the confluent hypergeometric function and

$$z = \text{sgn}(V_0)\sqrt{\sigma x} + \sqrt{2\rho}. \qquad (5)$$

However, this solution is not correct. This is immediately checked by direct substitution of function (4) into the Heun equation (3). An alternative simple checking can also be done, for instance, by putting, for simplicity, $c_2 = 0$, then expanding function (4) into power series at point $x_0 = 2\rho/\sigma$ and substituting the resulting expansion into the Heun equation. One is then readily convinced that the result is not zero, hence, the function (4) does not solve the Heun equation (3).

The correct solution of equation (4) is presented in [2]. It is not written as a linear combination of a Hermite function and a Kummer confluent hypergeometric function. Rather, the correct solution is given by the Clausen generalized hypergeometric functions $_3F_2$ as [2] (see also [3-5])

$$u = c_1 \cdot {}_3F_2\left(\beta, \gamma, 1+\frac{\beta\gamma}{k}; \frac{\beta\gamma}{k}, a; \frac{x+1}{2}\right) + c_2 \cdot {}_3F_2\left(\beta, \gamma, 1-\frac{\beta\gamma}{k}; -\frac{\beta\gamma}{k}, c; \frac{1-x}{2}\right), \quad (6)$$

where $c_1$ and $c_2$ are arbitrary constants. Note that the exponent parameters involved in equation (3) obey the Fuchsian condition $c = 1 + \beta + \gamma - a - b$ and the accessory parameter $k$ satisfies the equation

$$k^2 + k(a-c) - \beta\gamma = 0. \quad (7)$$

We emphasize that this solution applies if $b = -1$. This is why we have stressed above that the specification $a = -1$, claimed by the authors, is not correct.

**2.** The authors claim that the potential (1) supports infinitely many bound states with energy levels given as

$$E_n = n^{-2/3}\left(\frac{-mV_0}{\hbar^2}\right)^{1/3}\frac{V_0}{2}, \quad n = 1, 2, 3, \cdots \quad (8)$$

However, this is an incorrect assertion because the number of bound states is in fact finite as shown in [2]. This is because the potential is a short-range one with exponentially vanishing behavior as $x$ goes to infinity. It can readily be checked that the integral of the function $xV(x)$ over the positive semi-axis is finite; hence, according to Bargmann's test [6], the potential indeed supports only a finite number of bound states. A rather accurate estimate for the number of bound states can be derived by checking the integral by Calogero $\int_0^\infty \sqrt{-V(x)}dx$ [7] and further applying the asymptotic result by Chadan [8]. As a result, one obtains [2]

$$n \leq 2(\sqrt{2}-1)\sqrt{m\delta^2|V_0|/\hbar^2}. \quad (9)$$

**3.** The authors claim that the bound-state wave functions are constructed by polynomial reductions of the function (4) if one puts $c_2 = 0$:

$$\psi_n = \left(H_n(z) - \sqrt{2n} H_{n-1}(z)\right) e^{-\sqrt{2n}z - \sigma x/2}, \tag{10}$$

where $z = \sqrt{2n} - \sqrt{\sigma} x$. In the light of aforesaid, this is of course a wrong assertion. For instance, the ground state wave function constructed in this way is given as

$$\psi_1 = \left(1 - \sqrt{2}z\right) e^{-\sqrt{2}z - \sigma x/2}. \tag{11}$$

It is straightforward to check that this function does not satisfy the Schrödinger equation (2).

One should also note that the polynomial reductions (10) are such that they do not vanish at the origin. For instance, for the ground state $n = 1$ we have from equation (11)

$$\psi_1\big|_{x=0} = -e^{-2} \neq 0. \tag{12}$$

However, according to the detailed analysis presented, e.g., in [9], the bound state wave functions should vanish at the origin. This requirement leads to an energy spectrum which differs from (8) by a non-zero Maslov index (see [11]).

**4.** The correct bound-state wave functions are presented in [2]. As it was mentioned above, the functions are given in terms of the Clausen generalized hypergeometric functions $_3F_2$. Alternatively, these functions can be written in terms of the familiar Gauss hypergeometric functions $_2F_1$ as [2]

$$\psi = (z+1)^{\alpha_1}(z-1)^{\alpha_2} \left( {_2F_1}\left(\alpha, \beta; 1+2\alpha_2; \frac{1-z}{2}\right) + \frac{2\alpha_2(\alpha_1 - \alpha_2)}{\alpha\beta - 2\alpha_2(\alpha_1 - \alpha_2)} \, {_2F_1}\left(\alpha, \beta; 2\alpha_2; \frac{1-z}{2}\right)\right), \tag{13}$$

where $z = \sqrt{1 - e^{-y/\delta}}$, and

$$\alpha_1 = \sqrt{\frac{2m\delta^2}{\hbar^2}(-E - 2V_0)}, \quad \alpha_2 = \sqrt{\frac{-2m\delta^2 E}{\hbar^2}}, \tag{14}$$

$$\alpha, \beta = \alpha_1 + \alpha_2 \pm \sqrt{\frac{8m\delta^2}{\hbar^2}(-E - V_0)}. \tag{15}$$

The exact equation for energy spectrum is given as [2]

$$_3F_2\left(\alpha, \beta, 1 - \frac{\alpha\beta}{q}; -\frac{\alpha\beta}{q}, 1 + 2\alpha_2; \frac{1}{2}\right) = 0. \tag{16}$$

This equation can be conveniently rewritten in terms of the hypergeometric functions $_2F_1$ as

$$S(E) \equiv 1 - \frac{\alpha\beta + 2\alpha_2 q}{2\alpha_2 q} \frac{{_2F_1}(\alpha, \beta, 1 + 2\alpha_2; 1/2)}{{_2F_1}(\alpha, \beta, 2\alpha_2; 1/2)} = 0. \tag{17}$$

With this spectrum equation, it is straightforwardly checked that the bound-state wave functions (13) satisfy the Schrödinger equation (2) for potential (1). The normalized wave functions for parameters chosen as $m, \hbar, V_0, \delta = 1, 1, -4, 2$ are shown in Fig. 1. It is seen that the wave functions indeed vanish at the origin.

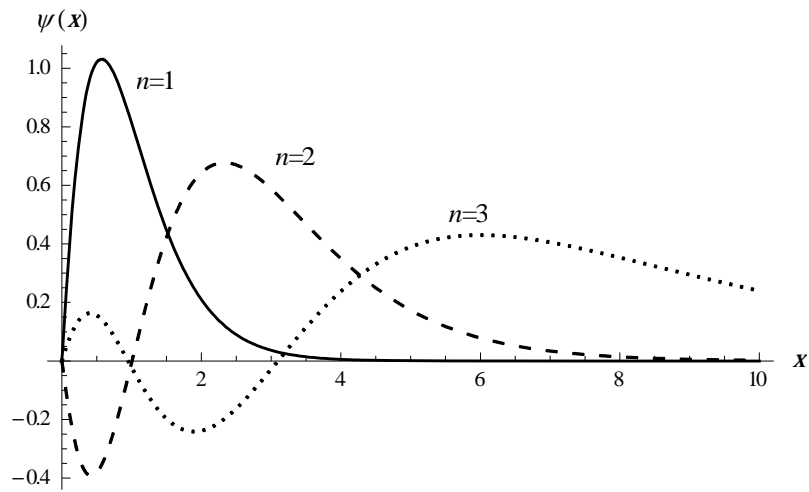

Fig. 1. Normalized wave functions (reproduction from [2]).

The plot of function $S(E)$ for parameters $m, \hbar, V_0, \delta = 1, 1, -4, 2$ is shown in Fig. 2. In this case the function has only three roots [2]:

$$E_{1,2,3} = -0.0294695, -0.4166327, -2.1680511, \qquad (18)$$

so that the number of bound states is three. We note that the estimate (9) for the number of bound states gives $n \leq 3.31$, which is indeed an accurate estimate.

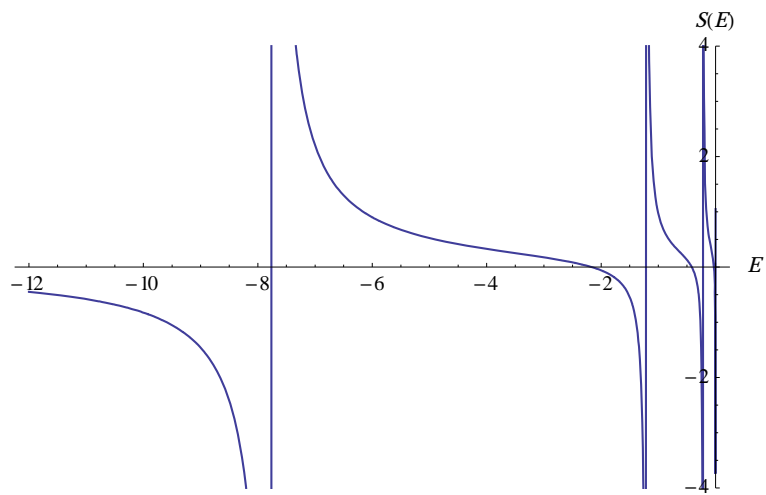

Fig.2. Graphical representation of the spectrum equation (17) (reproduction from [2]).

**5.** The authors of the commented paper discuss the third harmonic generation based on the transitions between discrete energy levels. We note that such a generation is possible only if three exist at least *four* discrete energy levels. Since the number of the bound-state energy levels is approximately proportional to $\sqrt{|V_0|}$ (see estimate (9)), an immediate conclusion is that the generation is possible only if the potential strength $|V_0|$ exceeds the minimal *threshold* estimated using inequality (9) as

$$|V_0| \geq \frac{4(3+2\sqrt{2})\hbar^2}{m\delta^2} \approx \frac{23}{\delta^2}\frac{\hbar^2}{m}. \tag{19}$$

For example, since there are only three discrete levels for parameters $m, \hbar, V_0, \delta = 1, 1, -4, 2$, it is understood that the third harmonic generation in this particular case is not possible at all.

To get a general insight on the possibility of harmonic generation (more precisely, the possibility of wave mixing such as sum-frequency generation, difference-frequency generation, etc.), it is helpful to look at the behavior of the bound-state energy levels, corresponding energy intervals and the transition matrix elements as functions of the potential parameter $V_0$. These dependences for $m, \hbar, \delta = 1, 1, 2$ are shown in Figs. 3,4 and 5. As seen from Fig. 3, for small $|V_0| < 1.3$ one has only one energy level, other levels gradually emerging as $|V_0|$ increases. A three-wave mixing (e.g., the second harmonic generation) becomes possible if the third level emerges at $|V_0| \approx 3$. A four-wave mixing (e.g., the third harmonic generation) is possible when the fourth level emerges. For the particular parameters chosen, this happens when $|V_0| > 5.5$.

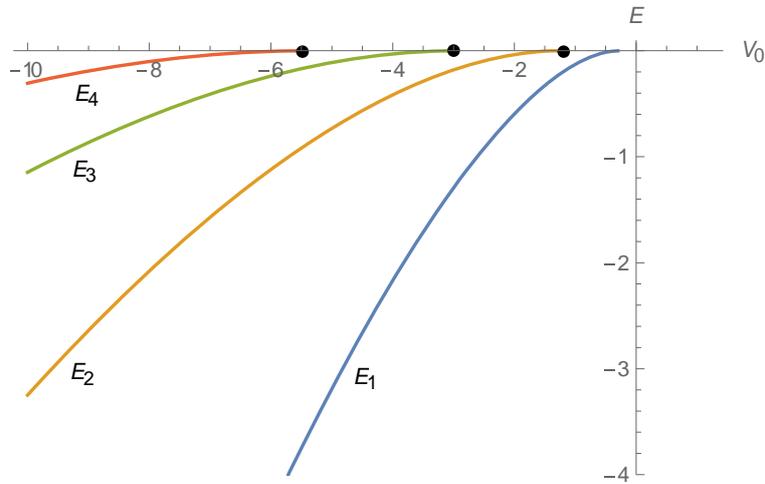

Fig. 3. Energy levels vs. $V_0$ if $m, \hbar, \delta = 1, 1, 2$. The filled circles indicate the points where a now energy level emerges.

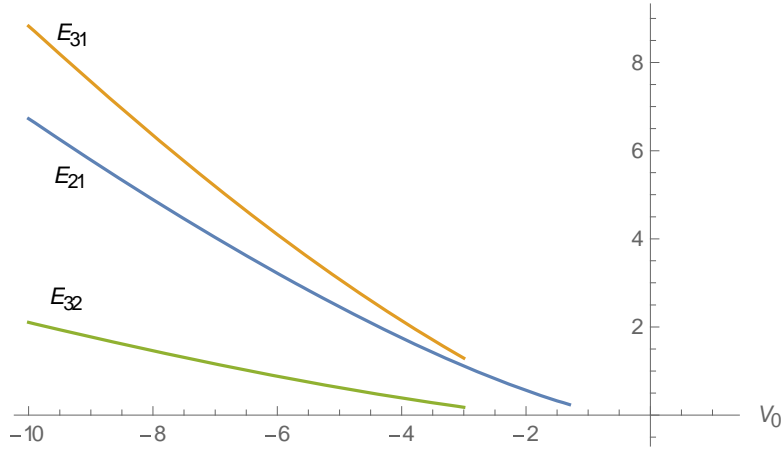

Fig. 4. Energy intervals vs. $V_0$ if $m, \hbar, \delta = 1, 1, 2$.

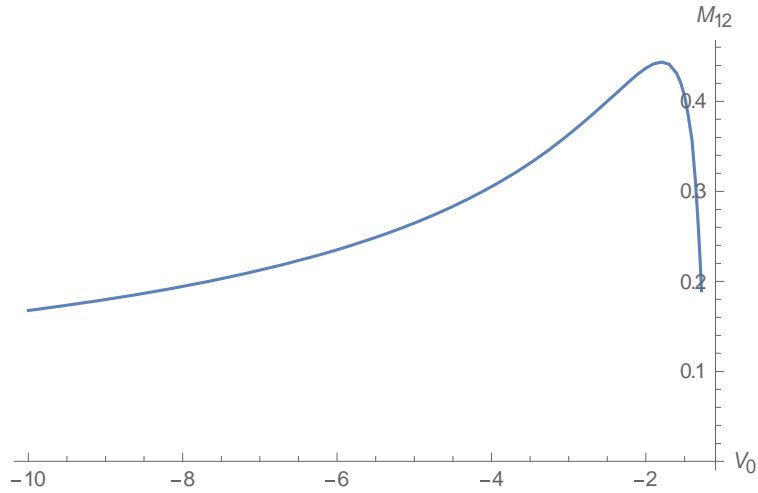

Fig. 5. Transition matrix element $M_{12}$ vs. $V_0$ if $m, \hbar, \delta = 1, 1, 2$.

**Discussion**

The authors consider the third harmonic generation by a quantum system bound in the short-range bottomless exponential potential well (1), which belongs to the general Heun classes of potentials [10]. The treatment of the paper is incorrect in several instances.

First, the correct solution of the Schrödinger equation for this potential is written in terms of the Clausen generalized hypergeometric functions [2] (or, alternatively, as an irreducible linear combination of two Gauss ordinary hypergeometric functions, see [5]), while the authors claimed a solution being a linear combination of a Hermite function and a Kummer confluent hypergeometric function.

Second, the authors claim that the number of bound states supported by the potential is infinite, while the potential is a short-range and thus supports only a finite number of bound states. This is readily checked by Bargmann's test [6].

Third, the bound state wave functions used in the paper are incorrect. It should be noted that the wave functions used by the authors refer in fact to the inverse-square-root potential treated in [11]. Note also that the latter potential belongs to a different Heun class. Namely, the inverse-square-root potential belongs to a bi-confluent Heun class [12-14].

Fourth, the bound state wave functions used in the paper do not vanish at the origin, while the well known analysis strongly indicates that they should vanish [9].

We hope that our contribution will serve the authors of the original article and the readers of the journal to improve their knowledge of physical systems such as those discussed here.

**Acknowledgments**

This research was supported by the Russian-Armenian (Slavonic) University at the expense of the Ministry of Education and Science of the Russian Federation, the Armenian Science Committee (SC Grant No. 18T-1C276), and the Armenian National Science and Education Fund (ANSEF Grant No. PS-5701).